\documentclass[aps,prb,showpacs,twocolumn,superscriptaddress,floatfix]{revtex4} 

\usepackage{graphicx}
\usepackage{bm}
\usepackage{epsfig}
\input epsf

\begin{document}

\newcommand{\nsl}[1]{\rlap{\hbox{$\mskip 1 mu /$}}#1}
\preprint{APS/123-QED}

\title{A Criterion for the Critical Number of Fermions and 
Chiral Symmetry Breaking in  Anisotropic $QED_{3}$}

\author{ A. Concha}
\author{V. Stanev }
\author{Z. {Te{\v s}anovi{\'c}}}

\affiliation{Department of Physics and Astronomy, Johns Hopkins University, Baltimore, Maryland 21218}


\begin{abstract}
By analyzing the strength of a photon-fermion coupling
using basic scattering processes we calculate the 
effect of a velocity anisotropy on the critical number of fermions at which mass is 
dynamically generated in planar QED. This gives a quantitative criterion which
can be used to locate a quantum critical point at which fermions are gapped and
confined out of the physical spectrum in a phase diagram
of various condensed matter systems. We also discuss the mechanism of relativity
restoration within the symmetric, quantum-critical phase of the 
theory.
 
\end{abstract}

\vskip 1cm
\maketitle

\section{Introduction}

Quantum Electrodynamics (QED), the quantum theory of radiation and its 
interaction with matter, is a subject with far reaching impact in physics, from the 
calculation of the magnetic moment of the electron 
\cite{Schwinger} to the  widespread use of diagrammatic tools in every 
corner of theoretical physics. However, despite this long
tradition as a central paradigm in physics, QED
continues to confront us with many new challenges
as its modern reincarnations emerge as effective theories of strongly
correlated many-body problems, from quantum spin liquids to high
temperature superconductivity to graphene.
%
Back in Feynman's day, we would have been very surprised if someone were 
trying to solve a relativistic problem with two speeds of light. These days, such
problems are actually ubiquitous in modern theoretical physics, one example being
the effective theory of low 
energy excitations in a correlated d-wave superconductor.
That theory is equivalent to an {\em anisotropic} quantum electrodynamics 
in $2+1$ dimensions\cite{franz2001,franz2002i}, in which different
``speeds of light'' appear naturally. 
Furthermore, there are other systems in
which the low energy description reduces to 
different versions $QED_{3}$, such as various forms of quantum
spin liquids \cite{hermele,saremi} or the physics of graphene layers \cite{herbutgraphene}.
All of these different problems share a common 
feature: they have a nodal structure that resembles a relativistic spectrum 
for the low energy excitations. However, a distinctive 
feature of the $QED_{3}$ particularly
relevant for cuprate superconductors is that it contains a significant intrinsic 
anisotropy, which exists due to the difference between the Fermi and gap 
velocities ($v_{F}$, $v_{\Delta}$).

It is known that the {\em fermionic} anisotropy, $\alpha=v_{F}/v_{\Delta}$, is irrelevant 
in the perturbative renormalization group (RG) sense\cite{vafek},
 as long as the system is in 
the {\em symmetric} phase of $QED_{3}$.
This is the quantum-critical phase of the theory, in which strongly interacting massless
fermions acquire anomalous power-law behaviors in their various correlation functions.
The anomalous dimension exponents characterizing this unusual state are
universal and typically depend
only on the total number of fermion flavors, $N$. This however is true only as long as 
$N>N_{c}$, where $N_{c}$ is the 
critical number of fermions at which the fermion  mass is dynamically generated via 
the mechanism of spontaneous chiral symmetry breaking (CSB). Once CSB takes place,
the fermions are gapped and confined out of the physical spectrum. This heralds 
a different, massive phase of the theory which typically translates to a different
state in the underlying condensed matter system. The results of Ref. 7
are generally valid for arbitrary anisotropy as long as the 
number of fermions $N$ is sufficiently large or for small anisotropy when
$N$ is greater than $N_c$ of the isotropic case. 

Evidently, 
$N_{c}$  is an important number within the theory, not in the least
because antiferromagnetic order in effective theories of high temperature superconductors and
quantum spin liquids generically arises via 
the above non-perturbative phenomenon of CSB -- for example, in the context 
of cuprates, the chiral mass generation corresponds 
to the onset of a spin density-wave order from
within a quantum disordered d-wave 
superconductor \cite{herbut,tesanovic}
while it describes the formation of the Ne\' el antiferromagnetic state
and a whole family of other order states in the context of quantum spin liquids.
\cite{hermele,saremi} 
Unlike the exponents of the critical massless phase, however,
$N_{c}$ itself is not universal.
Consequently, an important question arises within 
the anisotropic $QED_{3}$: to what extent 
is the critical 
number of fermions, $N_{c}$,  which defines the boundary 
between broken and unbroken chiral symmetry, affected when such anisotropy is present?

In this paper, our goal is to provide an answer to this question. 
Of course, ours is not {\em the} answer, for two reasons. First, since
$QED_{3}$ is a strongly interacting theory its exact behavior is beyond our reach. 
Second, since $N_{c}$ is not universal there are actually {\em many different} $N_{c}$'s:
the nominally irrelevant couplings within the $QED_{3}$ theory of
a quantum disordered d-wave superconductor\cite{herbut,tesanovic} are
very different than those of lattice-based quantum spin liquids.\cite{hermele,saremi}
Furthermore, both these $N_{c}$'s are different from the {\em intrinsic} 
$N_{c}$ of the pure $QED_{3}$ field theory considered here (defined through
Balaban-Jaffe regularization,\cite{balaban} for example).
However, all these issues notwithstanding,
even in the absence of the exact solution, it is still possible to make a 
rather accurate determination of the {\em parametric} dependence of $N_{c}$ on the
anisotropy, once an ``exact'' $N_{c}$ is known for the isotropic case from a different
source, say from numerical simulations. Our goal is to devise a criterion for
determination of $N_{c}$ within the anisotropic $QED_{3}$ which, while not exact, still
provides a rather accurate description of how the CSB boundary changes as a 
function of the parameters of the theory. The philosophy here is
similar to the one behind the ubiquitous Lindemann\cite{lindemann} criterion, originally
proposed to predict the melting point of a solid. While not exact,
the Lindemann criterion has proven itself a 
remarkably accurate and useful in a wide range of situations, from classical to quantum
solids, from Wigner crystals to Abrikosov vortex lattices.

To devise our criterion, we point 
out that mass generation is a consequence of the fermion-photon coupling strength in QED. 
This strength is generally renormalized from its bare value by virtual polarizability
of the vacuum. 
Relying on this fact, we propose a natural way to measure the strength of the gauge field 
by focusing on the matrix element that represents 
processes in which one photon is exchanged between two fermions.
We stipulate that the CSB and mass generation take place when this matrix element
exceeds certain critical value. Within the isotropic $QED_{3}$ this is manifestly
an exact statement -- the only unknown is the actual value of $N_{c}$ which we can 
either infer  from a separate calculation 
or borrow from numerical simulations.\cite{hands,thomas,strouthos}
Once our ``Lindemann criterion'' is calibrated in this fashion, 
we proceed to evaluate
the appropriate matrix element in the anisotropic theory 
and propose that the CSB takes place
when this matrix element exceeds the {\em same} critical value. 

Following the above procedure, our criterion allows us to derive the following
main results: first, we show that $N_{c}$ is not just a function of the 
anisotropy $\alpha $, but also a non-monotonic function of $v_{F}$ and $v_{\Delta}$
themselves. The reason behind 
this peculiar behavior is the existence of a third 
relevant velocity in the theory, the speed of light, $c_{s}$, which naturally appears
in the Maxwellian action for the gauge field. 
We confirm our findings for $v_{\Delta}>c_{s}$ by using the 
Schwinger-Dyson (SD) equation within the Pisarski approximation;\cite{pisarski}
both results show the same functional dependence on $v_{\Delta}$ 
in this regime. Another result is that the critical number of fermions decreases
even for $\alpha=1$, as long as $v_F=v_{\Delta}\gg c_{s}$.
Thus, in the context of the $QED_{3}$ effective theory of underdoped cuprates, 
we generically 
expect a non-superconducting, non-magnetic pseudogap state to intrude
prior to the emergence of 
antiferromagnetism as d-wave superconductor is underdoped toward half-filling.  

\section{Criterion and Calculation}

We start by defining the form of the anisotropic $QED_{3}$ Lagrangian and 
the main physical quantities that will be used in our 
calculations.
The Lagrangian is
\begin{eqnarray}
\mathcal{L}_{Aniso}=\sum_{l=1}^{N}\bar{\psi}_{l}v_{\mu}^{(l)}\gamma_{\mu}(i\partial_{\mu}-a_{\mu})\psi_{l}+
\mathcal{L}[a_{\mu}(x)]
\label{lagrangian}
\end{eqnarray}
where $ \bar{\psi}_{l} $ are four component spinors and 
$a_{\mu}$ is a gauge field. In condensed matter environment, Eq. (\ref{lagrangian}) 
is the low energy effective theory of some strongly correlated electron system.
For example, in a phase-disordered d-wave superconductor, the 
gauge field $a_{\mu}$ couples nodal Bogoliubov-deGennes-Dirac 
fermions to fluctuating quantum 
vortex-antivortex pairs through a Maxwellian action $\mathcal{L}[a_{\mu}(x)]$. 
$\gamma_{\mu}$ are the standard gamma matrices, such 
that $\{\gamma_{\nu},\gamma_{\mu}  \}=\delta_{\mu \nu}$, and $v_{\mu}^{(l)}$ is defined 
as $(1,v_{F},v_{\Delta})$ and $(1,v_{\Delta},v_{F})$ for nodes $l=1,2$ 
respectively. Fermi and ``gap'' velocities, $v_F$ and $v_{\Delta}$, define
fermionic anisotropy at each node.

To make further progress we first focus our attention on a 
particularly simple case: the non-interacting one 
($a_{\mu}=0$). This case is equivalent to saying 
that the phase is rigid or that the departure 
point in our analysis is the fully-ordered superconducting state.
In that case Eq. (\ref{lagrangian}) is reduced to
\begin{eqnarray}
\mathcal{L}_{Dirac}=\sum_{l=1}^{2}\bar{\psi}_{l}v_{\mu}^{(l)}\gamma^{\mu} (i \partial_{\mu}+m)\psi_{l}
\end{eqnarray}
 where  a mass term has been introduced only for normalization purposes. 
In the end our results 
will be robust in the $m\rightarrow 0$ limit.

The equations of motion for these non-interacting Dirac fermions are:
\begin{eqnarray}
(-v_{\mu}^{(l)}\gamma^{\mu}i\partial_{\mu}+m)\psi_{l}=0
\end{eqnarray}      
with $l=1,2$. From this expression it is 
clear that the task of solving Eq. (3) can be performed independently for 
each fermion flavor $1$ and $2$.
Now, if we focus on $l=1$, we can reduce the anisotropic 
equation of motion to the isotropic one using a 
simple change of variables $\tau'=\tau$, $x'=v_{F}x$, and $y'=v_{\Delta}y$. 
It is worth  remarking 
that the intrinsic fermionic anisotropy {\em cannot} be removed in the case of the
real d-wave superconductor  once the gauge field is present, 
as can be seen from the definition 
of $v_{\mu}^{(l)}$.

We now outline how a typical 
perturbative approach for computing the effect of coupling to the gauge field
works: starting from the solutions 
obtained for the problem in which 
$a_{\mu}=0$, namely $\psi_{1}(\vec{x})$ and $\psi_{2}(\vec{x})$ -- both having the anisotropy 
encoded in their spatial oscillations -- we will perturbatively introduce the field $a_{\mu}$.
First, we choose our free solutions for each node in the form
 $\psi^{+}_{l}(\vec{x})=s_{r}(m,0)\exp(-imt)$ 
and $\psi^{-}_{l}(\vec{x})=t_{r}(m,0)\exp(+imt)$, where for convenience we have 
defined $ s_{1}(m,0)=(1,0,0,0)$, $ s_{2}(m,0)=(0,1,0,0)$, $ 
t_{1}(m,0)=(0,0,1,0)$, $t_{2}(m,0)=(0,0,0,1)$. The eigenstates of Eq. (3) are then
\begin{equation}
\psi_{1}(x)=\sum_{\vec{p}\hspace{1pt}r=1,2}\left(\frac{m}{V E_{\vec{p}}}\right)^{\frac{1}{2}}\left(b_{r\hspace{2pt} \vec{p}}\bar{s}_{r}(\vec{p})e^{-i\vec{p}\cdot\vec{x}}+d_{r \hspace{2pt}\vec{p}}\bar{t}_{r}(\vec{p})e^{i\vec{p}\cdot\vec{x}}\right)
\end{equation}
where $b_{r\hspace{2pt} \vec{p}}$ and $d_{r \hspace{2pt}\vec{p}}$ are standard fermionic 
annihilation operators, while $\vec{p}$ and $r$ are the momentum and spinor indices, respectively.
In the case of decoupled nodes, when $a_{\mu}=0$, all the information about the 
anisotropy is encoded in the exponential factor. 
Now, for finite momentum we can use the ansatz  
$\psi^{+}_{1}(\vec{x})=s_{r}(\vec{k})\exp(-i\vec{k}\cdot \vec{x}')$ 
and $\psi^{-}_{1}(\vec{x})=t_{r}(\vec{k})\exp(+i\vec{k}\cdot \vec{x}')$, where 
$\vec{x}'=(\tau,v_{F}x,v_{\Delta}y)$ for node $1$. 
The spinors 
$s_{r}(\vec{k})$ and $t_{r}(\vec{k})$ can be found directly from the equation of motion
\begin{eqnarray}
(\nsl{k} -m)s_{r}(\vec{k})=0 && (\nsl{k}+m)t_{r}(\vec{k})=0
\end{eqnarray}
where $\nsl{k}=\gamma^{\mu}k_{\mu}$. 
Thus, applying  $(\nsl{k}+m)$ and $(\nsl{k}-m)$ to $s_{r}(m,0)$ 
and $t_{r}(m,0)$ respectively we can generate the 
needed solutions.

Using these free fermions in the Heisenberg picture, 
we now introduce the gauge field by employing the well-known 
formal solution of the scattering matrix $S$, equivalent to:
\begin{eqnarray}
S=T[\exp(-i\int d^{2+1}x \mathcal{H}_{I}(\vec{x}))]
\end{eqnarray}
where $T$ and $:$ are the time ordered operator and the normal order operation, respectively. The 
interaction Hamiltonian density that couples fermion fields to
the gauge field is given by
\begin{eqnarray}
\mathcal{H}_{I}(\vec{x})=\sum_{l=1}^{2}:\bar{\psi}_{l}v_{\mu}^{(l)}\gamma_{\mu}a_{\mu}\psi_{l}:
\end{eqnarray}
Now, as usual, we want to analyze the matrix element $<f|S|i>$, where both 
the final and initial state are
assumed to be eigenstates of the {\em unperturbed} Hamiltonian. 
\begin{figure}[th]
\leavevmode
\epsfxsize=8.5cm
\epsfbox{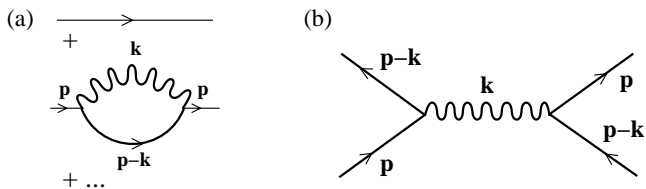} 
\caption{\label{Fig1}(a) Diagramatic form of the Schwinger-Dyson equation at one loop approximation.
(b) Diagram used to measure the strength of a photon.} 
\noindent
\end{figure}
Using the Heisenberg representation of the free 
fields $\psi(\vec{x})$ and the Wick's theorem, we recognize that 
the only relevant diagrams are the interactions photon-fermion and fermion-fermion, the 
so called direct scattering and the M$\ddot{o}$ller scattering. 
The calculation simplifies here because, 
due to its fluctuating nature, the contribution of 
$a_{\mu}$ to the first term of the $S$ matrix expansion 
forces the said matrix element to vanish. Consequently, the leading
contribution relevant for our purposes comes from  
the second order term where the contraction of $a_{\mu}(\vec{x}_{1})$ with 
$a_{\nu}(\vec{x}_{2})$ is nothing else but the photon propagator 
$D_{\mu \nu}(\vec{x}_{1}-\vec{x}_{2})$, which is the  real space {\em inverse} of 
the polarization function 
\begin{eqnarray}
\Pi_{\mu \nu}=\sum_{n}\frac{N}{16v_{F}v_{\Delta}}\sqrt{k_{\alpha}g^{\left(n\right)}_{\alpha \beta}k_{\beta}}\left(g^{\left(n\right)}_{\mu \nu}-\frac{g^{\left(n\right)}_{\mu \rho}k_{\rho}g^{\left(n\right)}_{\nu \lambda}k_{\lambda}}{k_{\alpha}g^{\left(n\right)}_{\alpha \beta}k_{\beta}}\right)
\end{eqnarray}
where the diagonal nodal {\em ``metric''}  $g_{\mu \nu}^{\left(n\right)}$ is defined\cite{vafek} by
$g^{\left(1\right)}_{00}=g^{\left(2\right)}_{00}=1$, $g^{\left(1\right)}_{11}=g^{\left(2\right)}_{22}=v_{F}^{2}$ and
 $g^{\left(1\right)}_{22}=g^{\left(2\right)}_{11}=v_{\Delta}^{2}$, and 
$\gamma_{\nu}^{\left(n\right)}=\sqrt{g^{\left(n\right)}_{\mu\nu}\gamma_{\mu}}$. 
We will now demonstrate that this second order element of the perturbation theory 
can be used to gainfully define the ``strength'' of the fermion-photon interaction
and determine the critical number of fermionic flavors $N_{c}(\alpha)$,
without ever solving the Schwinger-Dyson equation, even though a perturbative approach 
itself is condemned to failure.

In order to accomplish this, let us focus on the mass generation problem. 
It is well known that Schwinger-Dyson equation, Fig. 1(a), 
has a non trivial solution ($m>0$) once 
the initially soft photon (factor 
$1/N$ in the large $N$ limit) becomes harder at a critical number of fermions, $ N_{c}$;
 with $3<N_{c}<5$  \cite{pisarski,nash,appelquist,hands}.
 That means for  $N<N_{c}$ mass 
will be dynamically generated and therefore chiral symmetry would be broken.

The existence of anisotropy makes to analyzing the birth 
of a gapped state through the mechanism of mass generation using the Schwinger-Dyson 
formalism almost impossible, with the noteworthy exception of the isotropic limit and the so called 
small anisotropy case ($v_{F}=1+\delta$,$v_{\Delta}=1$, 
where $\delta \ll 1$) \cite{lee,vafek,stanev}.
Our aim is to explore mass generation for arbitrary anisotropy, because this is the physically relevant regime. 
Our technique is based on the observation that as we lower $N$ we are effectively making the photon
interaction stronger. This is clear because the photon propagator contains a screening factor $\frac{1}{N}$. From this simple observation we can 
see that mass generation is a phenomenon that is intrinsically 
tied to the strength of the gauge field (photons).
\begin{figure}[th]
\leavevmode
\epsfxsize=8.5cm
\epsfbox{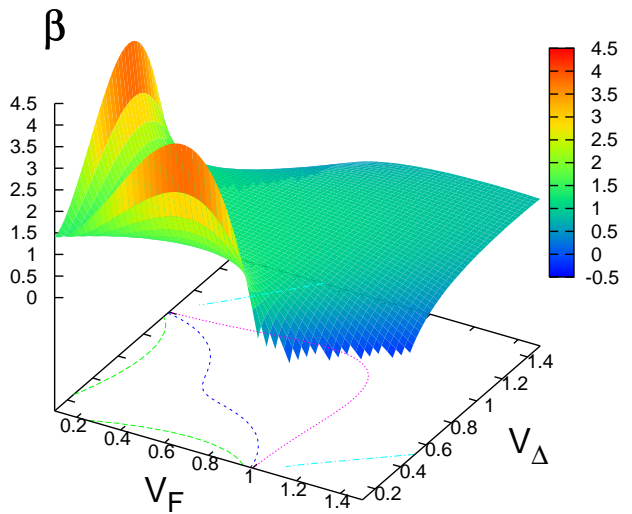} 
\caption{\label{Fig2}
 The calculated ratio $\beta$. The minimum value of $v_{F}$and 
$v_{\Delta}$ is $0.15$ and the maximum $1.5$. This graph shows that $N_{c}$ is not a simple 
function of $\alpha$ but a more complex function of $v_{F}$ and 
$v_{\Delta}$ .}
\noindent
\end{figure}
Put simply, we can translate the meaning of $N_{c}$
into the following statement, which clearly is a necessary condition for mass generation: 
{\em If the photon-exchange interaction is stronger than some threshold value}
$A_c$ {\em then the fermion mass will be dynamically generated}.
This assertion clearly lacks the precision of a mathematical theorem
but we now proceed to demonstrate that it has the practical virtue of a 
useful criterion.

The problem that we are facing is straightforward: How can we usefully 
quantify the $\it{strength}$ of the gauge field?
The simplest way to measure the strength of a photon is to analyze a process in which one 
photon is exchanged. For the sake of simplicity, we will analyze the $\it{fermion-fermion}$ 
scattering mediated by one $\it{photon}$, as shown in Fig. 1(b) which corresponds 
to the second order term in perturbation theory. 
Thus using the $N_{c}$ borrowed from the isotropic limit, we define 
 $A_{c}=<i|S^{(2)}|f>[N_{c}(\alpha=1)]$, the matrix element of the $S$ matrix evaluated at 
the isotropic limit, as the natural candidate to measure the strength of the photon. We can 
safely say that in an anisotropic theory mass will be generated at the critical strength 
defined by $A_{c}$.

 Thus solving the equation, $A_{c}=<i|S^{(2)}|f>[N_{c}(\alpha)]$ for $N_{c}(\alpha)$, we 
will find the 
influence of anisotropy on the critical number of fermion flavors.
In the isotropic case, the choice of  node will not make any difference, however in the 
anisotropic case the situation will be quite different. Regarding the symmetries 
of the problem we must analyze the 
invariant, $ <i|S^{(2)}|f>(v_{F},v_{\Delta})+<i|S^{(2)}|f>(v_{\Delta},v_{F}) $ because nodal particles can 
be born indifferently in nodes $1$ or $2$.
This quantity is invariant under the 
transformation $v_{F}\leftrightarrow v_{\Delta}$, which is 
a physical symmetry of this problem.

We now proceed to compute the first non zero matrix element $<i|S^{(2)}|f>$. The second order
process following from the S-matrix term, Eq. (6) is:
\begin{equation}
S^{(2)}\propto T[\int \mathcal{H}_{I}(\vec{x}_{1})\mathcal{H}_{I}(\vec{x}_{2})dx_{1}dx_{2}] 
\end{equation}
which can be written (integration with respect to $x_{1}$ and $x_{2}$ is implicit) as a sum of terms of the type   
\begin{equation}
T[:\bar{\psi}_{1}(x_{1})\gamma_{\mu}^{(1)}a_{\mu}(x_{1}) \psi_{1}(x_{1})::\bar{\psi}_{1}(x_{2})\gamma_{\nu}^{(1)}a_{\nu}(x_{2}) \psi_{1}(x_{2}):]
\end{equation}  
By applying Wick's theorem to the previous expression terms with and
without contractions are obtained. However, due to the fluctuating nature 
of the gauge field most of those terms will vanish. The survivor will 
be the one containing the contraction of the gauge field with itself. Thus, the 
problem is reduced to compute the matrix element:
\begin{equation}
:\bar{\psi}_{1}(x_{1})\gamma_{\mu}^{(1)} \psi_{1}(x_{1})\bar{\psi}_{1}(x_{2})\gamma_{\nu}^{(1)} \psi_{1}(x_{2}) D_{\mu \nu}(x_{2}-x_{1}):
\end{equation}
To evaluate such matrix element we must define the initial and final states of the system
$|i>,|f>$. As the strength of the photon should be independent of the states that we use to 
measure it, we are free to use 
\begin{equation}
{|i>=b^{\dagger}_{r_{1} \vec{p}_{1}} b^{\dagger}_{r_{2} \vec{p}_{2}}|0> } \text{ and }
{|f>=b^{\dagger}_{r'_{1} \vec{p'}_{1}} b^{\dagger}_{r'_{2} \vec{p'}_2}|0>}\nonumber
\end{equation}
which we already know from the free theory.
The relevant matrix element is
\begin{widetext}
\begin{eqnarray}
\frac{1}{N}<i|S^{(2)}|f>(\Lambda,v_{F},v_{\Delta}) \sim\frac{1}{v_{\Delta}v_{F}} 
\int d^{3}k
\left(\frac{1}{p_{i0}(p_{i0}-k_{0})}\right)
Q_{\mu}(\vec{p}_{i}-\vec{k},\vec{p}_{i})D_{\mu \nu}(g^{(1)}\vec{k}) Q_{\nu}(\vec{p}_{i},\vec{p}_{i}-\vec{k})
\label{matrix}
\end{eqnarray} 
\end{widetext}
 where $Q_{\mu}(\vec{P},\vec{p})=\bar{s}(\vec{P})\gamma^{(1)}_{\mu}s(\vec{p})$, with $\bar{s}=s\gamma_{0}$.  
The integral is to be performed over all the allowed $k-space$, but we have one constraint that is the 
spectrum of the Dirac fermions, $k_{0}=\sqrt{\left(v_{F}k_{x}\right)^{2}+\left(v_{\Delta}k_{y}\right)^{2}}$, which 
allows us to further reduce the evaluation of this integral. 

The integration limits are fixed from the condition that the momentum transferred in the
processes shown in Fig. 1(b) can not be larger that the momentum of the incident particle, $\vec{p}_{i}$.
We have set $p_{x}=p_{y}=\Lambda$ as an upper cut off. We have also verified that the only effect of 
choosing a different cut off for different spatial directions is that it makes the convergence slower.
Thus, using the invariant measure of the strength of the gauge field, it is straightforward to 
show that the ratio 
between critical number of fermions at velocities $v_{F}$ and $v_{\Delta}$ respect to the 
isotropic point is 
\begin{widetext}
\begin{equation}
\beta=\frac{<i|S^{(2)}|f>(\Lambda,v_{F},v_{\Delta})+<i|S^{(2)}|f>(\Lambda,v_{\Delta},v_{F})}{2<i|S^{(2)}|f>(\Lambda,1,1) } 
\end{equation}
\end{widetext}
Thus, the critical number of fermions for the anisotropic theory is given by the 
product of, $\beta(v_{F},v_{\Delta})$, and  the 
critical number of fermions at the isotropic point $N_{c}\left(1,1\right)$.
In principle, this measure of the strength of the 
gauge field depends on the initial momentum of 
the particles, but it turns out that the ratio between 
the strength of the gauge field with anisotropy 
and  the strength of the gauge field without anisotropy $\alpha=1$ is not sensitive to the momentum 
of the incident fermions, $\vec{p}_{i}$. In fact, all the 
calculated quantities converge to a fixed value 
for a fairly low momentum cut off, $\Lambda$. It is important to remark that $\beta$ is gauge 
invariant due to the transverse nature of the photon. 

Indeed we can do better. By isolating the leading 
divergent behavior of the scattering amplitude we have 
obtained that in the $\Lambda \rightarrow \infty$ at the leading order in the 
large $N$ approximation, the ratio $\beta$ is given by:
\begin{widetext}
\begin{eqnarray}
\beta\left(v_{F},v_{\Delta}\right)&=&\zeta_{1}
\left[
\sqrt{1+v_{\Delta}^{2}}\mathbf{E}\left(\frac{v_{\Delta}^2- v_{F}^{2}}{1+v_{\Delta}^2}\right)+
\sqrt{1+v_{F}^{2}}\mathbf{E}\left(\frac{v_{F}^{2}-v_{\Delta }^2}{1+v_{F}^2}\right)
\right]+
\zeta_{2}
\left[
 \mathbf{K}\left(\frac{v_F^2-v_{\Delta }^2 }{1+v_{F }^2}\right)+
 \mathbf{K}\left(\frac{v_{\Delta }^2-v_{F}^{2} }{1+v_{\Delta}^2}\right)
\right]
\label{result}
\end{eqnarray} 
\end{widetext}
where
\begin{eqnarray}
&&\mathbf{E}\left(m\right)=\int_{0}^{\frac{\pi}{2}} \left[1-m \sin(\theta ) \right]^{1/2} d \theta \nonumber\\
&&\mathbf{K}\left(m\right)=\int_{0}^{\frac{\pi}{2}} \left[1-m \sin(\theta ) \right]^{-1/2} d \theta \\
&&\zeta_{1}= 
-2\left(\frac{ 1 -2\left(v_{F}^2+ v_{\Delta }^2\right)- v_{F}^{2}v_{\Delta }^2 +v_{F}^4+ v_{\Delta }^4 }{3\sqrt{2}
\pi v_{F}^{2}v_{\Delta}^{2}}\right)\nonumber \\
&&\zeta_{2}=
\frac{2-3\left( v_{F}^{2}+v_{\Delta}^{2}\right)+ \left(v_{F}^{2}-v_{\Delta}^{2}\right)^{2}+
v_{F}^{4}v_{\Delta}^2+ v_{\Delta}^{4}v_{F}^{2}}{3\sqrt{2}
\pi v_{F}^{2}v_{\Delta}^{2}\sqrt{\left(1+v_{F }^2\right)\left(1+v_{\Delta}^2\right)}}\nonumber
\end{eqnarray}
are the elliptic functions of first and second kind respectively,
and the prefactors are polynomials that are symmetric under $v_{F}\leftrightarrow v_{\Delta}$.
Thus $\beta$ is explicitly invariant under the transformation  $v_{F}\leftrightarrow v_{\Delta}$, as it must be.

An important case is the one in which there is no
 $\it{fermionic}$ $\it{anisotropy}$, {\em i.e.}
 $v_{F}=v_{\Delta}$. In this case all the integrals can 
be performed and the full  divergent behavior 
of Eq. (\ref{matrix}) can be isolated by going to cylindrical 
coordinates as the $x\leftrightarrow y$ symmetry suggests.
The result is just the limit behavior  of Eq. (\ref{result})  when taking $v_{F}=v_{\Delta}$. 
In that limit the dependence on the fermionic velocities becomes particularly simple.
\begin{eqnarray}
\beta=\sqrt{\frac{2}{1+v_{\Delta}^{2}}}
\label{beta}
\end{eqnarray} 
To test the accuracy of our technique, the result is compared with the exact numerical 
integration in Fig. 3, showing excellent agreement. For 
low speeds $v_{\Delta}<1$, our result suggest that the critical number of fermions increases. 
We discuss this case in more detail below. 
On the opposite limit when the gauge field velocity is small compared with the other two velocities the critical 
number of fermions goes to zero.  
\begin{figure}[th]
\leavevmode
\epsfxsize=8.0cm
\epsfbox{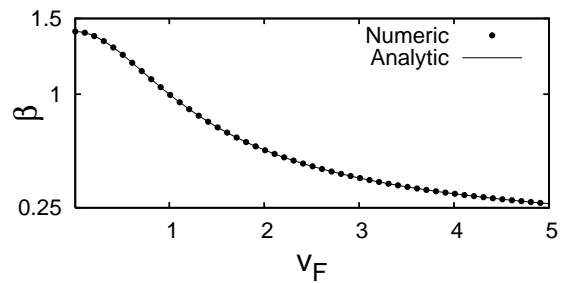} 
\caption{\label{Fig3}
Comparison between the numerical and analytical calculation of $\beta$ in the case for which $v_{F}=v_{\Delta}$.}
\noindent
\end{figure}

A non trivial dependence on the 
gauge field velocity was also found; there are three 
different regimes depending upon the value of $v_{F}$,  
with respect to $c_{s}$, as shown in Fig. 2.
For $v_{F}<c_{s}$, the critical number of fermions increases, 
reaching unexpected values for large 
anisotropy, leading to a gapped state in this sector. On the other 
hand for  $v_{F}>c_{s}$, $N_{c}$ decreases
and we can generically expect gapless excitations. 
The most unexpected result occurs if $v_{F}=c_{s}$, in 
this case $N_{c}(\alpha)\approx N_{c}(1)$ for $\alpha > 1$. 
Even though we found small deviations
with respect to the isotropic value near to $\alpha=2$, 
they do not change the integer part 
of $N_{c}$ (See also Fig. 5). Thus, in this regime large 
anisotropy will be completely irrelevant.
\begin{figure}[th]
\leavevmode
\epsfxsize=7.5cm
\epsfbox{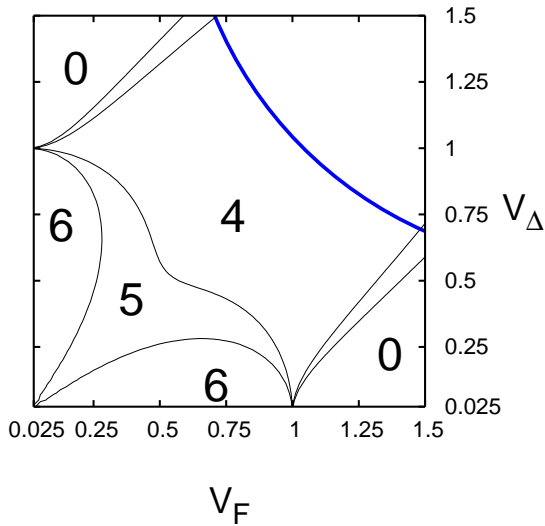} 
\caption{\label{Fig4}Phase diagram for the critical number of fermions in 
the velocity space. We 
have set the critical number of fermions for the 
isotropic theory at $N_c=4$. Inside the square 
defined by the lines $v_{F}=1$ and $v_{\Delta}=1$ the critical 
number of fermions never vanishes. The 
blue line corresponds to $v_{F}=1/v_{\Delta}$ and represents the one dimensional domain 
simulated by Thomas and Hands\cite{hands}.}
\noindent
\end{figure}
   
Using $N_{c}(1)=4$, which is  the integer closest to the 
gauge invariant critical number of fermions $128/3\pi^2$ found by Nash \cite{nash}, we have 
obtained a phase 
diagram in the ($v_{F}$, $v_{\Delta}$) space, Fig. 4. 
In this plot it is clear that if both velocities 
are smaller than $c_{s}$, the critical number of fermions 
never goes to zero. However, outside that 
square region $N_{c}\rightarrow 0$ for high anisotropies.

To show how important it is to consider the scale defined by $c_{s}$ we have also 
plotted  $\beta(\alpha)$ for different values of $v_{F}$  ($0.5,1,5.0$) that we kept fixed 
as $v_{\Delta}$ was varied (See Fig. 5). In this plot we make evident that $c_{s}$ is a relevant parameter 
which we are allowed to set to $1$, however the detailed behavior of the system is not a simple 
function of the anisotropy, $\alpha$, but an explicit function of both parameters $v_{F}$ and $v_{\Delta}$.

In order to test our results we compare them with recent numerical work that has been done 
by Thomas and Hands (T-H)\cite{hands} in an attempt to 
understand how anisotropy can modify the properties of $QED_{3}$.
The lattice (Euclidean) version used by T-H includes an anisotropy which is intended 
to mimic the continuum model in the incarnation presented by 
Lee and Herbut \cite{lee}. The $QED_{3}$ 
theory presented in Lee's article is parametrized in terms of two 
quantities $\delta=\sqrt{v_{F}v_{\Delta}}$ and  $\lambda=v_{F}/v_{\Delta}$.
T-H have used an extended lattice model similar to 
the one used by Dagotto et al \cite{dagotto}, that in the continuum limit resembles
 the behavior of Eq. (1). In order to perform the simulation the action used was:
\begin{eqnarray}
S=\sum_{i=1}^{N}\sum_{x,x'}a^{3}\bar{\chi}_{i}(x)M_{x,x'}\chi_{i}(x')+
\frac{\beta}{2}a^{3}\Theta^{2}_{\mu \nu}(x)
\end{eqnarray}
where the anisotropy was introduced in the fermion matrix 
\begin{equation}
M_{x,x'}=\frac{1}{2a}\sum_{\mu=1}^{3}\xi_{\mu}(x)
\left[\delta_{x',x+\hat{\mu}}U_{x\mu}-\delta_{x',x-\hat{\mu}}U^{\dagger}_{x'\mu}  \right]+m\delta_{\mu \nu}
\end{equation}
and
\begin{eqnarray}
\xi_{\mu}(x)&=&\lambda_{\mu}\eta_{\mu}(x)\\
\eta_{\mu}(x)&=&(-1)^{x_{1}+...+x_{\mu}}
\end{eqnarray}
with $x_{1}=x$, $x_{2}=y$ and $x_{3}=\tau$, is the Kawomoto-Smit phase of the staggered fermion field.
The lattice spacing is $a$. An important definition is that of the anisotropy factors, 
$\lambda_{x}=\alpha^{-\frac{1}{2}}$, $\lambda_{y}=\alpha^{\frac{1}{2}}$ , and $\lambda_{\tau}=1$ . This 
definition is important for our purposes because it shows 
that T-H lattice theory does not keep the flavor
 symmetry of the model relevant for cuprates\cite{franz2001,vafek,franz2002i,hands}. 
Regardless this intrinsic drawback of the method we will show that 
it still mimic the qualitative behavior of the flavor 
symmetric anisotropic $QED_{3}$ at least 
in the low anisotropy limit.

In their numerical simulation T-H have a single parameter, which is equivalent 
to $\alpha=v_{F}/v_{\Delta}$, and as an extra constraint they have set $\delta=1$. This 
choice imply that $v_{F}=1/v_{\Delta}$ and therefore $\alpha=1/v_{\Delta}^{2}$.
 That means that T-H have simulated a 1-D domain of the whole parameter 
space $(v_{F},v_{\Delta})$. That domain is shown in Fig. 4 as a blue line.
In order to make a link between the amount of condensate $<\bar{\psi}\psi>$ and 
its relation with $N_{c}(\alpha)$ we will assume that the functional form of the dynamically 
generated mass does not change \cite{appelquist,nash} as we introduce anisotropy in the system.
This can be explicitly checked in the small anisotropy limit\cite{stanev}.
As long as we are inside to the broken phase and near to the boundary 
between broken and unbroken phases, the mass has the following functional form
\begin{eqnarray}
m=m_{o}\exp\left( \frac{-2\pi}{\sqrt{ \frac{N_{c}(v_{F},v_{\Delta})}{N}-1}}\right)
\label{mass}
\end{eqnarray}
where $m_{o}$ may depend on $N$. However, what is important is how to define the boundary between 
the massless phase and the massive phase. This is done by solving Eq. (\ref{mass}) when $m=0$. We can also 
interpret this equation in the 
following way: For a fixed number of Fermions, $N$, this 
equation allow us to understand how the mass changes as a function of $N_{c}(\alpha)$ when 
$N$ is close to $N_{c}$.
In fact, for any number of fermions when $N_{c}(\alpha)=N$ it is prohibited to have any condensate. 
That means that the critical anisotropy $\alpha_{c}$ that solve $m(\alpha_{c})=0$ for any value of $N$ is 
the same that the critical anisotropy that solves $N_{c}(\alpha_{c})=N$. Hands reported 
that such a critical anisotropy $\alpha_{c}\approx 4$ in a $16^{3}$ sites lattice 
simulation which is in agreement
with the critical value found by us $\alpha_{c}\approx 3$. This 
decreasing behavior is in contrast 
to the one found in Ref. 18, where it was claimed that $N_{c}$ increases as a function of 
the bare anisotropy. Taking into account that the lattice simulation is a 
non perturbative method that does not relay in any educated ansatz, T-H results strongly 
support our view of the phenomenon.   
Still, we should mention that the critical anisotropy calculated by T-H using the anisotropic scaling 
is not, in the strict sense, quantitatively accurate 
in the context of cuprates, for two reasons. First, 
the scaling used brakes the crystal isotropy, or, in different words, their simulation is not 
invariant under flavor exchange. Second, they assumed that the two fermionic velocities 
change little  around the gauge field velocity $c_{s}=1$. Due to the collective nature of 
phase defects we expect that $c_{s}\ll v_{F}$ in cuprates \cite{note1}.

On the other hand lattice simulations are always performed in finite lattices and therefore 
the correct comparison with our results should be done by considering both, an upper cut off 
and a lower cut off. In principle we are free to make the upper cutoff as large as we want but the lower 
cut off dependence may be important when comparing our results with finite lattice 
simulations\cite{gusynin}. Setting the upper cut off as $\Lambda_{u}=1$ then 
the lower cut off should be $\Lambda_{d}\sim 1/L$, where $L$ is the size 
of the system. As we change the lower cut off we have found no 
significant differences in our results as it goes to zero.   
\begin{figure}[th]
\leavevmode
\epsfxsize=8.5cm
\epsfbox{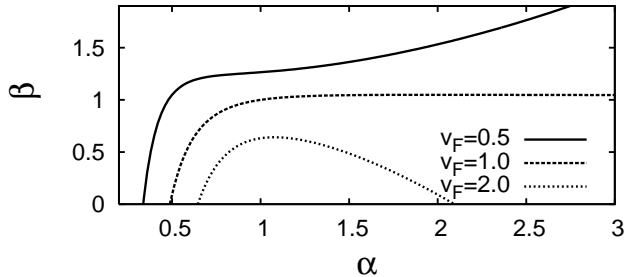}
\caption{\label{Fig5:}Ratio between the critical number of fermions in the anisotropic 
theory and its equivalent within the isotropic theory as a function 
of anisotropy, $\beta(\alpha)$. The 
three curves represent the three regimes found: $N_{c}(\alpha)<N_{c}(1)$ if $v_{F}>1$, 
$N_{c}(\alpha)/N_{c}(1)$ approaches a constant $\sim 1$ if $v_{F}=1$, 
and  $N_{c}(\alpha)$ increases for increasing $ \alpha $ if $v_{F}<1$.}
\noindent
\end{figure}
We have also investigated the effect of the breaking of the flavor symmetry in our scheme. For 
this purpose we have used only one amplitude for the photon, so that the final kernel is not 
invariant under $v_{F}\leftrightarrow v_{\Delta}$. In this case, we have obtained the same 
qualitative behavior as shown in Fig. 6, and the small negative slope at the isotropic point, detail 
that is in agreement with the results shown by T-H in [Fig. 5, (Ref. 11)]. This clearly is a 
symptom of the breaking of flavor symmetry. However, there is a very important issue that 
emerges once we arbitrarily break flavor symmetry. The function $\beta$ starts to pick up 
a phase which is unphysical. $\beta$ must be a real number as $N_{c} $ is. This makes evident 
that even with the simplest  interaction between fermions the flavor symmetry is needed in 
order to obtain a meaningful value of $\beta$. This observation suggest that  numerical 
simulations that preserve flavor symmetry are the only reliable way to extract accurate 
critical values. However, we must encore that the qualitative behavior of T-H simulation 
agrees with the physical picture proposed in this article.
\begin{figure}[th]
\leavevmode
\epsfxsize=7.5cm
\epsfbox{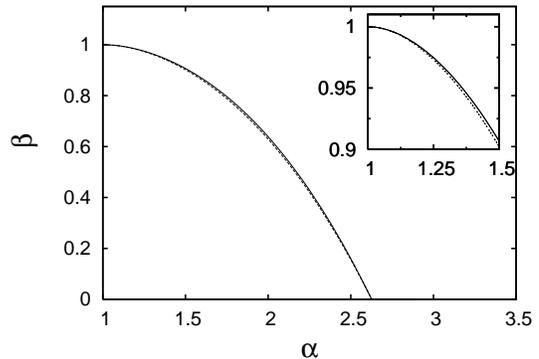}
\caption{\label{Fig6}$\beta$ within the domain defined by the 
blue line of Fig. 4. The continuous line 
represents the flavor symmetric calculation. The dotted line
is the non-symmetric calculation. For the latter case
we have only plotted the real part because the breaking of the flavor symmetry 
introduces a small but persistent imaginary part.}
\noindent
\end{figure}
Looking at the obtained phase diagram, Fig. 4, the question that
naturally arises is, which regime is 
the physically relevant one? Clearly our answer will depend on the ratio $v_{F}/c_{s}$.
At finite but still small temperature, $T$, we can use the continuum vortex-antivortex Coulomb 
plasma model in order to 
estimate the gauge field velocity. Identifying the speed of light from the Maxwellian form of the 
action for the gauge field we find that $c_{s}\sim \sqrt{n_{l}/T}$ at finite $T$, where $n_{l}$ is the 
density of vortices \cite{franz2002i}. 
Thus for any $T>0$ as we approach the
 superconducting state $n_{l}\rightarrow 0$ and 
thus $c_{s}\rightarrow 0$, resulting in a protected symmetric phase.
As $n_{l}$ increases, for small but finite $T$, $c_{s}$  may reach very high values, but if those 
values are larger than $v_{F}$  is unclear. If we go to the line $T=0$ quantum fluctuations will 
drive the system into a region in which the value of $c_{s}$ will depend on the specific value of the
dynamical critical exponent $z$, that in some simplified calculations was adopted to be $1$.
The more striking problem about identifying the precise value of $c_{s}$ is that this 
velocity is a function of the correlation length, $c_{s}=f(\xi_{sc})$,
but at the same time we know that
$\xi_{sc}=\xi_{sc}(x,T)$, thus a self consistent treatment or knowledge of
the correlation length from experiments will be needed to settle this problem 
and give an accurate phase diagram that identify the relevance of these different regimes.

To provide further evidence of our findings we have also 
re-analyzed from a different perspective 
the case in which $\it{fermionic}$ $\it{anisotropy}$ is one. We have applied 
Pisarski's technique \cite{pisarski} 
to find the qualitative behavior of the dynamically 
generated mass as a function of $v_{\Delta}$.

We will assume that $\Sigma\left(p\right)\sim \Sigma\left(0\right)$. Thus in the appropriate integration
interval $(\Sigma\left(0\right),\Lambda)$ the mass will be a constant. This assumption is certainly 
incorrect, as was shown by Appelquist \cite{appelquist,nash} et al. However, it will allow us 
to compare the qualitative behavior of the 
mass as a function of $v_{\Delta}$, for the case in which $v_{F}=v_{\Delta}$. In this case the sums
can be performed analytically by going to cylindrical coordinates instead of 
spherical ones. The integration was performed 
over the shell defined by a lower cutoff $\Sigma\left(0\right)$ and an upper cut off $\Lambda$. 
Thus the Schwinger-Dyson equation:
\begin{eqnarray}
\Sigma_{(n)}\left(\vec{p}\right)=
\int \frac{\gamma_{\mu}^{(n)}D_{\mu \nu}(\vec{p}-\vec{k})\gamma^{(n)}_{\nu}\Sigma_{(n)}(\vec{k})}
          {k_{\mu}g^{n}_{\mu \nu}k_{\nu}+\Sigma^{2}_{(n)}(\vec{k})}d^{3}k
\end{eqnarray} 
can be solved in this rough approximation and the result is:
\begin{eqnarray}
\Sigma(0)\sim \exp \left( -\frac{ N \pi^2 v_{\Delta}^{2}}{8\sqrt{v_{\Delta}^{2}-1}}\right)
\end{eqnarray}
which is a real number as long as $v_{\Delta}>1$. Thus as long as the $\it{neutrinos}$ move faster than 
the photons the generated mass indeed does depend on $v_{\Delta}$. 

We expect that -- given the fact the decay factor in Pisarski's 
result is of order $N_{c}$ -- the correction found
for that factor will give us the functional behavior of $N_{c}(v_{\Delta})$. 
Thus, as $v_{\Delta}\gg 1$ we 
would expect that $N_{c}\sim 1/v_{\Delta}$. This is consistent 
with the result obtained from the proposed criterion, Eq. (\ref{beta}). 
However, we should warn the reader that 
even thought a mathematical expression can be obtained for $v_{\Delta}<1$ 
the nature of the system 
will change in this case, casting doubt on
the validity of our criterion in that region. Indeed, from S-D equation at the 
Pisarski level approximation for low fermionic velocities, the self energy will acquire 
an imaginary part which can be interpreted as
leading to a confinement\cite{maris} for fermions of the theory. This instantly calls into
question  the validity of using a pure plane wave type
solutions for the computation of the scattering process.
  
In this approximation, the nature of the solution changes considerably 
at the point where $v_{\Delta}=1$. That is so because the radial integral 
gets a logarithmic contribution that is proportional to 
$\sqrt{1-v_{\Delta}^2}$ which overwhelms the leading contribution 
at the isotropic point. Alternatively, in more physical terms, if 
photons and massless fermions 
move at precisely the same speed this is ``infinitely'' different than having
the photons that move 
faster than fermions. In the latter case it is natural to expect an ``overscreening''
behavior, in which constant exchange of fast photons
ultimately leads to confinement. On the opposite side, with fermions moving faster
than photons, we expect photons to be less effective in screening the fermions, and thus 
less effective in generating their mass. We have also checked that in the isotropic limit 
the Pisarski's answer obtains and thus our results are not an artifact of the 
parametrization used.

\begin{figure}[th]
\leavevmode
\epsfxsize=8.3cm
\epsfbox{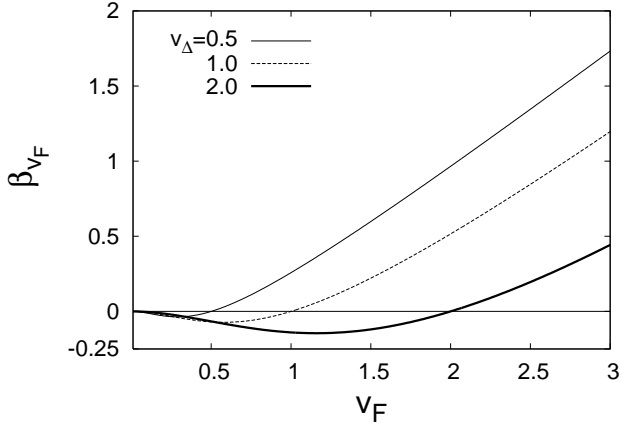}
\caption{\label{Fig7} The RG-$\beta$ function for $v_{F}$ (in arbitrary units) computed using the 
same method as in Ref. 7 but without setting $v_{\Delta}=c_{s}=1$. This plot shows the parametric 
dependence of the fixed point on the value at which $v_{\Delta}$ is initially set in this calculation.}
\noindent
\label{bfunction}
\end{figure}

\begin{figure}[th]
\leavevmode
\epsfxsize=7.0cm
\epsfbox{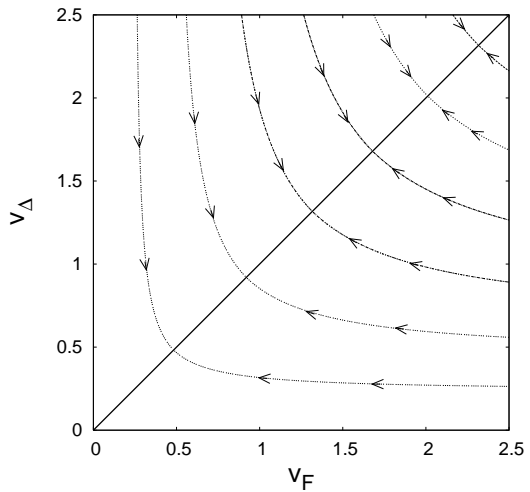}
\caption{\label{Fig8} The corresponding RG-flow of $v_{F}$ and $v_{\Delta}$ when charge renormalization 
is not taken into account. A line of fixed points at $v_{F}=v_{\Delta}$ is apparent.}
\noindent
\label{rgflow}
\end{figure}

So far, we have shown that mass generation has a non-universal behavior, which arise due to 
the breaking of Lorentz invariance. Thus, it is natural for a cautious reader to wonder 
if, once $N>N_{c}$, the renormalized effective
low energy theory is in fact Lorentz invariant or not. To 
begin with we emphasize that the fact that $\alpha$ flows to 
one by itself does not guarantee  that the full Lorentz invariance will emerge 
unless both $c_{s}$ and $v_{F}$ are set equal; this in effect acts as an
extra constraint. This is shown in Fig. (\ref{bfunction}) and in 
Fig. (\ref{rgflow}) where it is easy to see that even thought 
$\alpha \rightarrow 1$ neither $v_{F}$ nor $v_{\Delta}$ converge 
 to $c_{s}=1$ unless the above mentioned extra constaint is imposed.  To show that 
the full Lorentz invariance is indeed restored we must prove that 
both fermionic velocities $v_{F}$ and $v_{\Delta}$
flow to $c_{s}=1$ independently.

It is easy to see that Lorentz invariance can still be 
broken even with the fermionic anisotropy set to unity. Let us 
assume that the bare values of the Fermi and gap 
velocity are equal to each other but different from
the gauge field velocity, i.e.
$v_{F}=v_{\Delta}$, but $v_{F}\neq c_{s}$. The Lagrangian for this 
simplified theory is \cite{kaul,note2} 
\begin{widetext}
\begin{eqnarray}
\mathcal{L}=\bar{\psi}_{1}\left\{\gamma_{0}(\partial_{\tau}+ia_{\tau})+ 
\gamma_{1}v_{F}(\partial_{x}+ia_{x})+\gamma_{2}v_{F}(\partial_{y}+ia_{y}) \right\}\psi_{1}+
 (1 \leftrightarrow 2)+\nonumber\\
\frac{1}{2 e^2}
\left \{  \left( \frac{\partial a_{x}}{\partial y}-\frac{\partial a_{y}}{\partial x} \right )^{2}+
 \left( \frac{\partial a_{\tau}}{\partial x}-\frac{\partial a_{x}}{\partial \tau} \right )^{2}+
\left( \frac{\partial a_{\tau}}{\partial y}-\frac{\partial a_{y}}{\partial \tau} \right )^{2}  \right\}
\end{eqnarray}
\end{widetext}
By simple rescaling $\tau'=\tau$, $x=v_{F} x'$, $y=v_{F} y'$, $a'_{\tau'}=a_{\tau}$, 
 $a'_{x'}=v_{F}a_{x}$, and $a'_{y'}=v_{F}a_{y}$ we 
can transform this theory into
a new 
theory in which the fermionic part of the action remains fully 
isotropic but an anisotropic Maxwellian term appears:
\begin{widetext}
\begin{eqnarray}
\frac{1}{2 e^2 v^{4}_{F}}
 \left( \frac{\partial a'_{x'}}{\partial y'}-\frac{\partial a'_{y'}}{\partial x'} \right )^{2}+
\frac{1}{2 e^2 v^{2}_{F}} 
\left \{ \left( \frac{\partial a'_{\tau'}}{\partial x'}-\frac{\partial a'_{x'}}{\partial \tau'} \right )^{2}+
\left( \frac{\partial a'_{\tau'}}{\partial y'}-\frac{\partial a'_{y}}{\partial \tau'} \right )^{2}  \right\}~.
\end{eqnarray}
\end{widetext}

Thus, the effect of having $v_{F}\neq c_{s}$ reduces to 
anisotropic couplings in the Maxwellian self-action of the gauge field. 
We denote these couplings 
as $e^{2}_{\tau}\equiv e^{2}v^{4}_{F} \text{ and }e^{2}_{\perp}\equiv e^{2}v^{2}_{F}$. 
These two couplings can be interpreted as two anisotropic charges. 
Such anisotropic charges can
change the value of the critical number of fermions in the original 
theory, as already shown in Eq. (24). 
In contrast, within the isotropic $QED_{3}$,  the critical number of fermions does not 
depend on $e^{2}$. This is true as long as we have only one coupling constant, but once
we introduce two different couplings  this pleasing
behavior is lost.
This simple example shows that we can restore Lorentz invariance in the 
fermionic part of the action at the expense of
breaking the Lorentz invariance of the Maxwellian part.

To summarize, the above discussion shows that at a bare level 
there are two intrinsic anisotropies in this 
problem; $\alpha$ is the fermionic one, while
$\theta=\left(e_{\tau}/e_{\perp}\right) $ is the Maxwellian one. 
$e_{\tau}$ and $e_{\perp}$ are the couplings in the temporal  
and  spatial directions respectively.
Thus at a bare level of the $QED_{3}$ theory relevant for the 
cuprates there are four coupling 
constants $v_{F}$, $v_{\Delta}$, $e_{\tau}$, and $e_{\perp}$. We will now
show that, even though they contain anisotropy both
in the fermionic and Maxwellian terms, in the large $N$ limit the relativity 
is ultimately restored, without any assumptions about the size of the anisotropy.

To make good on the above claim notice that the 
effect of fermions on photons is still described by 
Eq. (8). However, the gauge field stiffness is now:
\begin{eqnarray}
\Pi^{(0)}_{\mu \nu}=
\frac{1}{2e_{\tau}^{2}}\bar{\epsilon}_{\chi \delta \mu}\bar{\epsilon}_{\chi \lambda \nu}k_{\lambda}k_{\delta}
\end{eqnarray}
where the anisotropic Levi-Civita symbol is defined as $\bar{\epsilon}_{\tau \eta \mu}=\epsilon_{\tau \eta \mu}$,
 $\bar{\epsilon}_{x \eta \mu}=\theta\epsilon_{x \eta \mu}$, 
$\bar{\epsilon}_{y \eta \mu}=\theta\epsilon_{y \eta \mu}$. Thus, the effective Lagrangian of the theory
can be written as:
\begin{eqnarray}
\mathcal{L}_{Aniso}=\frac{1}{2}\left( \Pi^{(0)}_{\mu \nu}+ \Pi_{\mu \nu} \right)a_{\mu}(k)a_{\nu}(-k)
\end{eqnarray}
where the effect of fermions has been introduced through
the polarization function.
 This expression 
allow us to find the renormalized couplings, comparing the original bare gauge field 
stiffness with the screened one:
\begin{eqnarray}
\Pi^{(R)}_{\mu \nu}= \Pi^{(0)}_{\mu \nu}+ \Pi_{\mu \nu} 
\end{eqnarray}
 We find that to the lowest order in $1/N$ the renormalized couplings are:
\begin{eqnarray}
\left(\frac{1}{e_{\tau}^{2}}\right)_{R}&=&
\left(\frac{1}{e_{\tau}^{2}}\right)+\frac{N}{16\bar{k}}v_{F}v_{\Delta}\\
\left(\frac{1}{e_{\perp}^{2}}\right)_{R}&=&
\left(\frac{1}{e_{\perp}^{2}}\right)+
\frac{N}{16}\left(\frac{1}{k^{(1)}}\frac{v_{F}}{v_{\Delta}}+ \frac{1}{k^{(2)}}\frac{v_{\Delta}}{v_{F}}\right)
\end{eqnarray}  
where $1/\bar{k}=1/k^{(1)}+1/k^{(2)}$, and  $k^{(i)}=\sqrt{ k_{\alpha} g^{(i)}_{\alpha \beta} k_{\beta}}$.

The above one loop renormalization of the anisotropic charges allows us
to set up the renormalization group (RG) equations for the beta-functions describing
the flow of different couplings. These equations are rather complicated and
we have been able to fully 
solve them only numerically. However, the following result is rather
simple and can be extracted in an analytic form: on general grounds we 
expect that the non trivial infra-red fixed point 
should remain once we introduce the
anisotropy, even though its position in parameter space may change.
In order to find the value of the renormalized couplings at the fixed point, we 
analyze the {\em difference} between the renormalized 
couplings of the Maxwellian action
\begin{eqnarray}
\left(\frac{1}{e_{\tau}^{2}}- \frac{1}{e_{\perp}^{2}}\right)_{R} =
\left(\frac{1}{e_{\tau}^{2}} - \frac{1}{e_{\perp}^{2}}\right)+\hspace{1.0in}\\ \nonumber
\frac{N}{16}\left[ \frac{1}{k^{(1)}}\left(v_{F}v_{\Delta}-\frac{v_{F}}{v_{\Delta}} \right)+
\frac{1}{k^{(2)}}\left(v_{\Delta}v_{F}-\frac{v_{\Delta}}{v_{F}} \right)\right]
\end{eqnarray}
If we now rearrange our RG equations so as to focus on the beta-function
for this difference between the renormalized charges, $\beta_{e_\tau -e_\perp}$,
the above equation implies that
\begin{eqnarray}
\beta_{e_\tau-e_\perp}\sim \frac{N}{16}\left[\frac{\left(v_{F}v_{\Delta}-\frac{v_{F}}{v_{\Delta}} \right)}{k^{(1)}}+
                            \frac{\left(v_{\Delta}v_{F}-\frac{v_{\Delta}}{v_{F}} \right)}{k^{(2)}}\right]
+\nonumber \\ 
+~(\cdots) \hspace{1.10in}~,
\end{eqnarray}
where $(\cdots)$ can be rewritten in terms of beta-functions
for all other couplings and thus must vanish at the putative fixed point.
Clearly, noting that $N$ is an arbitrarily large number, 
and that $k^{(i)}$ share the same sign, it follows that: 
\begin{eqnarray}
v^{(R)}_{F}v^{(R)}_{\Delta}-\frac{v^{(R)}_{F}}{v^{(R)}_{\Delta}}=0
\end{eqnarray}
from where $v^{(R)}_{\Delta}=1$, and, given that the theory is fully
invariant under the exchange $v_{F}\leftrightarrow v_{\Delta}$, 
it follows that $v^{(R)}_{F}=1$.
Putting this information back in the flow equations 
it is clear that $1/e_{\tau}^{2}$ and $1/e_{\perp}^{2}$ themselves
diverge with the same slope at the fixed 
point, and therefore their ratio $\theta=(e_{\tau}/e_{\perp}) \rightarrow 1$. 
This result shows that 
Lorentz invariance is restored and thus the previous 
results\cite{vafek,hermele,saremi} remain valid. However, we 
have made it clear that the physics behind the restoration of full
Lorentz invariance follows a path more subtle than previously explored: 
at the infra-red fixed point the relativity is 
restored due to the {\em interplay} between the velocity and 
charge renormalizations, the velocity renormalization by itself being 
insufficient to fully restore relativity of the theory. 

\section{Conclusions}
We have proposed a simple criterion that allows 
an explicit computation of the critical number of 
fermions $N_c$ in a theory that contains intrinsic 
anisotropies. We have checked, at the Pisarski's level approximation, that 
this criterion captures the functional dependence of $N_{c}(v_{\Delta})\sim 1/v_{\Delta}$ 
in the case in which 
an explicit expression can be obtained from analytic calculations.

Our criterion  suggests that lattice simulations should be 
performed in such a way that important
symmetries of the theory, namely  $v_{F}\leftrightarrow v_{\Delta}$, are 
protected. Otherwise, there is a danger of obtaining 
spurious results. In lattice $QED_{3}$, it seems worthwhile to investigate the existence 
of a possible confined phase in the region of the parameter space 
where fermionic velocities are small 
compared with the gauge field velocity. 
Another venue that remains to be explored, is the possible usefulness of 
similar criteria for the analysis of 
CSB or other non perturbative phenomena in other physical 
systems that also feature anisotropic couplings. 

Finally, we have shown that the velocity anisotropy in $(2+1)QED$ does affect the number 
of critical fermion flavors $N_c$ 
at which chiral symmetry is broken due to the phenomenon of mass generation,
even though the large $N$ theory remains fully relativistic in its critical phase. 
Surprisingly, $N_{c}$ is a  non-monotonic 
function of $v_{F}$ and $v_{\Delta}$, and, depending on 
the specific value of the ratio $v_{F}/c_{s}$, different regimes emerge. 
Our results imply that if phase 
fluctuations destroy the superconducting order in underdoped cuprates, we should expect a 
protected chirally symmetric critical phase -- i.e. the pseudogap within this
theory -- as doping decreases, before we reach the 
antiferromagnetic region in the phase diagram. Details of how will this 
sequence take place
depend on the specific value of the gauge field velocity $c_{s}$ for different compounds. 
We hope that our results will contribute to better understanding of 
the quantitative issues that surround the value of $N_c$ in various effective
theories and  motivate further research on the anisotropic 
incarnations of the $QED_3$ theory. 

\section{Acknowledgments}
We thank T. Senthil for useful comments. This work was supported in part 
by the NSF grant DMR-0531159.


\begin{thebibliography}{25}
\bibitem[{not()}]{note2}
\bibinfo{note}{In Ref.[24] the authors claim that fermionic anisotropy goes to
  one, and thus they obtain the simplified model we used to show that this can
  be thought as a model in which there are anisotropic couplings in the
  Maxwellian term. Given that the number of fermions in physical systems is
  modest it would be interesting to analyze the consequences of having such
  anisotropic couplings.}
\end{thebibliography}
\end{document}